\documentclass[%
 reprint,
 amsmath,amssymb,
 aps, prl
]{revtex4-2}

\usepackage{graphicx}
\usepackage{dcolumn}
\usepackage{bm}
\usepackage{xcolor}
\usepackage[mathlines]{lineno}
\usepackage{pdfpages}
\usepackage{pgffor}
\usepackage{soul}
\makeatletter
\AtBeginDocument{\let\LS@rot\@undefined}
\makeatother

\begin{document}

\title{Nonlocal Structural Effects of Water on DNA Homology Recognition}
\author{Ehud Haimov}\thanks{These two authors share equal contribution}
\author{Jonathan G. Hedley}\thanks{These two authors share equal contribution}
\author{Alexei A. Kornyshev}
 \email{Corresponding author: a.kornyshev@imperial.ac.uk}
\affiliation{
Department of Chemistry, Faculty of Natural Sciences, Imperial College London, Molecular Sciences Research Hub, White City Campus, Wood Lane, W12 0BZ, United Kingdom
}

\begin{abstract}
\textit{Abstract - }The mechanism behind mutual recognition of homologous DNA sequences prior to genetic recombination is one of the remaining puzzles in molecular biology. Leading models of homology recognition, based on classical electrostatics, neglect the short-range nonlocal screening effects arising from structured water around DNA, and hence may only provide insight for relatively large separations between interacting DNAs. We elucidate the role of the effects of the nonlocal dielectric response of water on DNA-DNA interaction and show that these can dramatically enhance the driving force for recognition.
\end{abstract}

\maketitle

\textit{Introduction - } The ability of homologous (identical/similar) DNA sequences to recognise each other in the complex environment of the cell \cite{zickler2006, barzel2008} plays a central role in both the exchange of parental genes in germ cells and in repairing DNA damage \cite{leach1996, aguilera2007}. In contrast to the Watson-Crick dogma, stating that DNA segments can only interact when they unzip, we now know that homolog DNA pairs may recognise each other at short distances without unzipping, sensing each other through electrostatic interactions of correlated ion charges mediated by the electrolyte. Such an effect is believed to originate from the sequence-dependent twist angle between adjacent base pairs in the double stranded DNA (dsDNA) helix. As homologous pairs of dsDNA will have identical/similar sequences, homologous structural motifs will be characterised by a highly correlated base pair rise and twist angle throughout their lengths. Homologous interaction can result from either reduced DNA-DNA repulsion between these homologous structural motifs or even attraction between them, depending on the amount and the distribution of condensed counterions close to the DNA surface, as predicted by theory \cite{kornyshev2001, kornyshev2007}. Over the last two decades, the mechanism of the initial step of homology recognition between bare DNA with adsorbed or condensed counterions, in the absence of any proteins, has been further explored (for review see \cite{kornyshev2010}) and substantiated by both theory \cite{kornyshev2001, kornyshev2009} and experiments \cite{wynveen2008, kornyshev2013, inoue2007, baldwin2008, danilowicz2009, olee2016}. In all existing theories, however, a uniform electrolyte environment was assumed even at close separation distances, effectively ignoring the possible complex effects of structured water forming around the DNA surface. Such effects, of course, must emerge from molecular dynamics simulations, but are not easy to observe, due to either the limited size of the simulation box, or the degree to which the system, including the water molecules, is coarse-grained \cite{yoo2016, luan2008, cortini2017}. We present here the first theoretical analysis of such effects.  

\textit{The model - } Consider two parallel dsDNAs immersed in an electrolyte solution (see Fig. 1a). Each molecule is characterised by a hydrophobic cylindrical core of radius $b \approx 6\text{ \AA}$ surrounded by two negatively charged helical phosphate strands at radius $a \approx 10\text{ \AA}$. Condensed counterions specifically adsorb along the major and minor grooves, reducing the intrinsic negative charge of DNA. Within the Debye-Bjerrum approximation, we treat the condensed counterion distribution as an effectively infinitessimal narrow shell at approximately the same radius as the phosphate strands. 

Spatial correlations of dipolar polarization fluctuations in a polar solvent lead to the nonlocal dielectric screening response of the electrolyte solution~\cite{kornyshev1981}. Here we focus on the orientational component of the total polarization, which is characterised by the longest correlation length $\Lambda$. Following Ref.~\cite{kornyshev1983}, we consider a Lorentzian approximation for the solvent’s longitudinal nonlocal wavenumber-dependent dielectric function, $\varepsilon_s(k)$, given by:
\begin{equation}
    \varepsilon_s(k)=\varepsilon_* + \frac{\varepsilon-\varepsilon_*}{1+\frac{\varepsilon}{\varepsilon_*}\Lambda^2k^2}
\end{equation}
where $\varepsilon$ is the macroscopic bulk dielectric constant of the solvent and $\varepsilon_*$ is the dielectric constant corresponding to higher frequency degrees of freedom of the solvent molecules, sometimes called the short-range dielectric constant. This simple Lorentzian form of the dielectric response does not take into account more complicated effects associated with density oscillations and overscreening, but it is known that such effects are smeared by the finite size of the charge groups and thermal fluctuations of the DNA molecule, thus using this form is a good first approximation. 

\begin{figure}[t!]
    \centering
    \includegraphics[width=\linewidth]{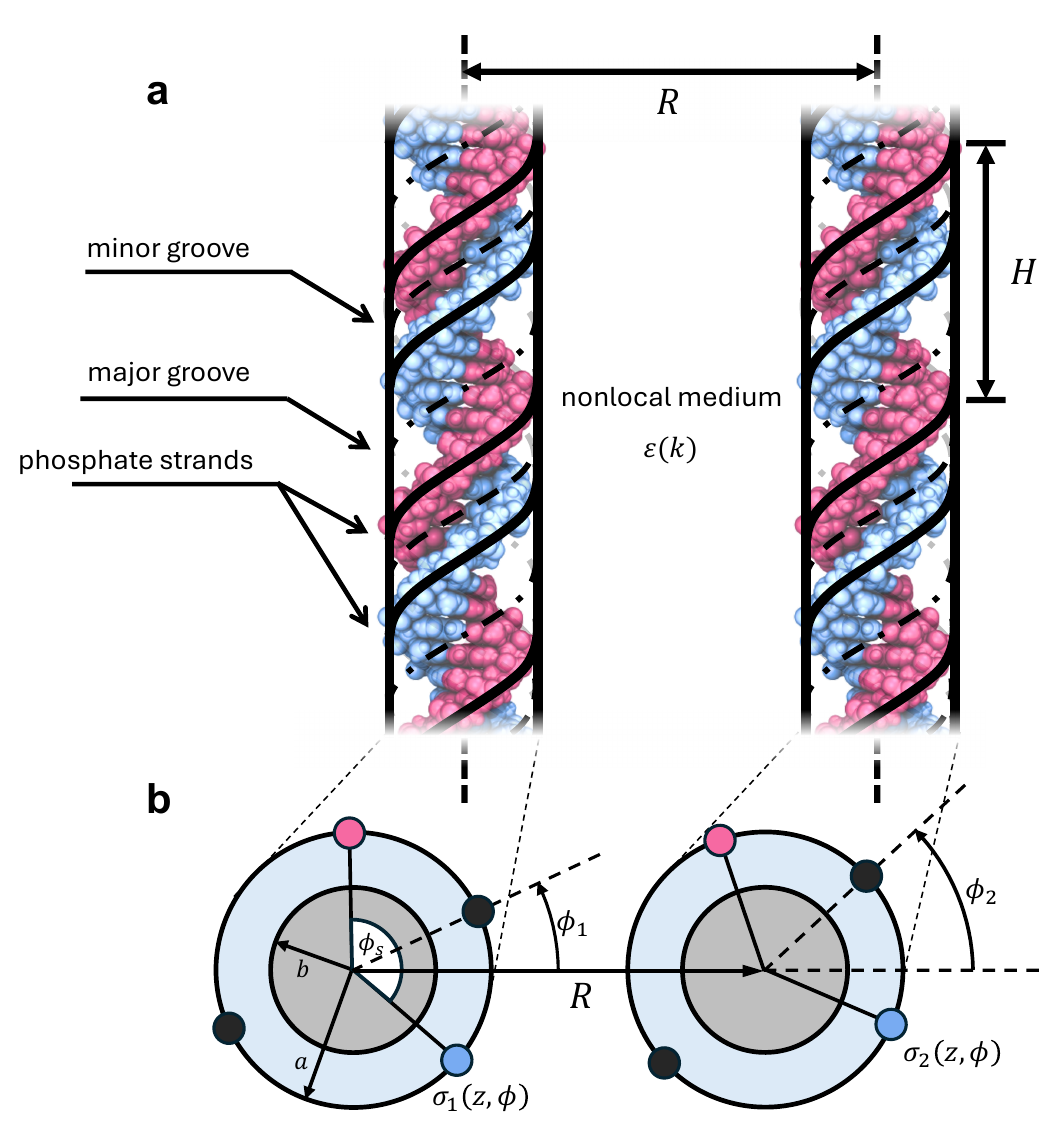}
    \caption{\footnotesize\textbf{Interaction of two homologous DNA molecules.} (a) Double helical structure of DNA molecule is modelled as two charged helices with pitch $H$ ($=34\text{ \AA}$ for B-DNA), with ions condensed in the major and minor grooves. The two interacting molecules are submerged in an electrolytic solution with nonlocal solvent dielectric function $\varepsilon(k)$, and separated by a interaxial displacement vector, $\mathbf{R}$. (b) Cross-sectional representation of the molecules. Each molecule has a cylindrical hydrophobic core with dielectric constant $\varepsilon_c$ and radius $b$, surrounded by a surface charge distribution $\sigma(z,\phi)$ at a radius $a$, describing the helical phosphate strands (pink and blue) and adsorbed counterions (black). The angle between phosphates is $\phi_s\approx0.8\pi$, describing the minor groove width. The coordinate $\phi_i$ is defined as the angle between $\mathbf{R}$ and the centre of the minor groove of molecule $i$.}
    \label{fig:1}
\end{figure}

Coupling the nonlocal solvent response to ion-ion correlations in aqueous electrolyte solutions is a tricky problem that is currently receiving a lot of attention \cite{kjellander2019, kjellander2016, becker2023}. The difficulty in modelling such an interplay at physiological concentrations arises from competing solvent-solvent and ion-ion correlation effects; at low (millimolar) ion concentrations, solvent correlations will dominate, however in highly concentrated solutions (e.g. solvent-in-salt systems) ion-ion correlations, beyond the Debye mean-field, will take control. Microscopic models of the nonlocal dielectric response of ionic systems have been proposed, leading to the modification of $\varepsilon_s(k)$, as well as a wavenumber-dependent inverse Debye length $\kappa(k)$ \cite{kjellander2019, kjellander2016, varela1998, budkov2020}. This procedure however is not straightforward, and does not result in any closed-form expression. Here we use a rudimentary formula for the dielectric function of the whole solution \cite{kornyshev1983},
\begin{equation}
    \varepsilon(k) =\varepsilon_s(k)+\frac{\kappa^2}{k^2}.
\end{equation}
where $\kappa^{-1}$ is the macroscopic inverse Debye length. This expression interpolates between low and high wavenumber limits. At small $k$, when $\varepsilon_s(k)\to\varepsilon$, the second term diverges as $\sim k^{-2}$, thus altogether yielding classical Debye screening behaviour. On the contrary, at large $k$, the second term vanishes and the dielectric response is determined by the solvent's $\varepsilon_s(k)$. This interpolation formula becomes exact in the limit of small electrolyte concentrations where $\kappa^{-1}$ is much smaller than any characteristic solvent correlation length. Within the Debye-Bjerrum approximation, taking condensed counterions into account, we adopt a linear response approach which can accommodate any form of $\varepsilon(k)$. This allows us direct comparison to models based on local approximations of the solution \cite{kornyshev1997, kornyshev1999}.

We calculate the electrostatic interaction energy for two distinct cases of parallel dsDNA pairs: ideal homologs (identical sequences) and non-homologs (uncorrelated sequences). The molecules are considered rigid, as the length along which the DNAs interact is taken as a single Kuhn length. This length is long enough to exhibit a significant difference between the interaction energies of homologs and non-homologs, but also short enough such that any kind of elastic effects can be ignored. The effects of torsional adjustment have been considered in the past  \cite{kornyshev2010, cherstvy2004, kornyshev2004} and can be introduced separately into the model. When considering longer interaction lengths, these effects can alter the recognition energy due to the `smiling effect’ ~\cite{kornyshev2010} and supercoiling~\cite{cortini2011}.

\begin{figure*}[t!]
    \centering
    \includegraphics[width=\linewidth]{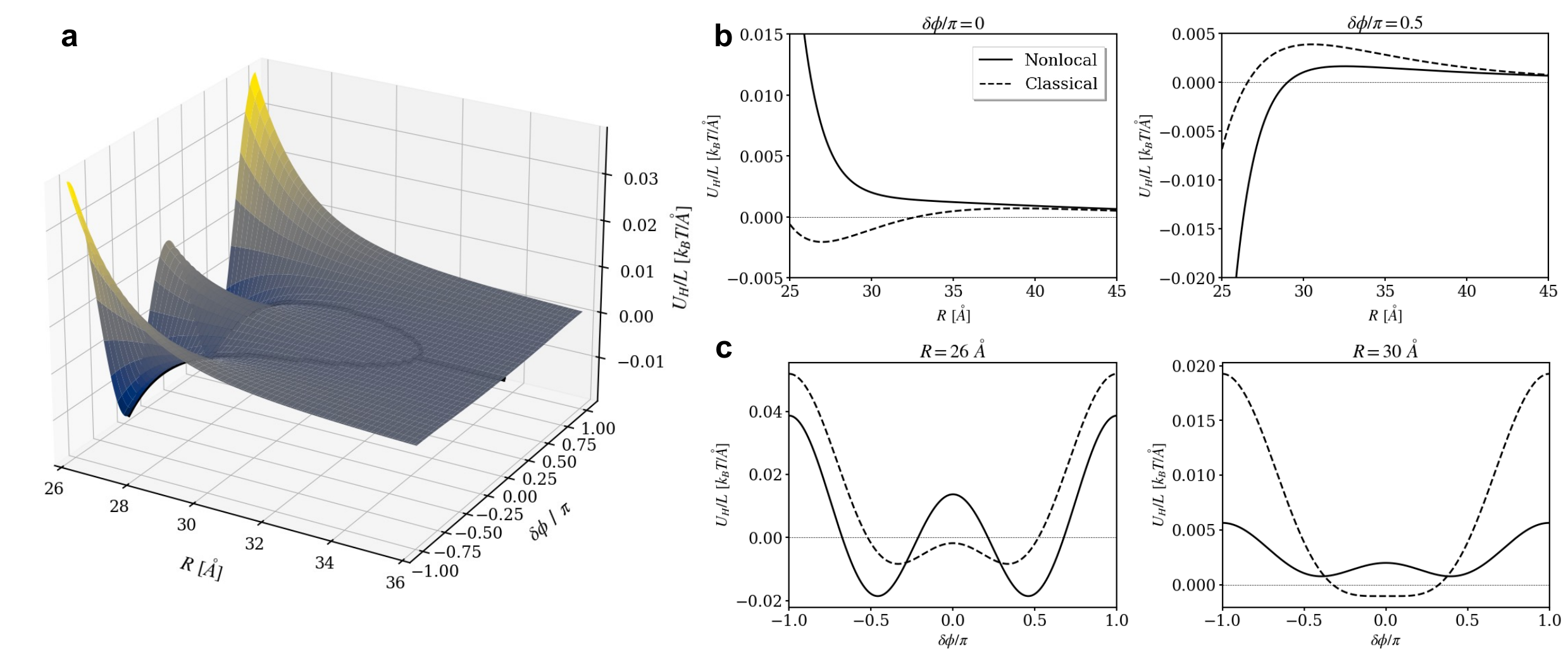}
    \caption{\footnotesize\textbf{Nonlocal electrostatic interaction of homologous dsDNA molecules.} For all figures, the DNA structural parameters are $g=2\pi/H$, $H=34 \text{ \AA}$, $h_r=3.4 \text{ \AA}$, $\phi_s=0.8\pi$, $\varepsilon_c=2$, $a=10 \text{ \AA}$ and $b=6 \text{ \AA}$. The counterion condensation distribution parameters are $f_1=0.3,f_2=0.7,f_3=0$, and $\Theta=0.8$. The nonlocal water parameters (see Eq. (1)) are $\varepsilon=80,\varepsilon_*=5,\gamma=\varepsilon/\varepsilon_* =16,\Lambda=3 \text{ \AA}$, and the Debye length for physiological concentrations ($\sim 0.154$ M) is calculated as $\kappa^{-1}=7.69 \text{ \AA}$. (a) Electrostatic interaction energy between DNAs as a function of interaxial separation (R) and the relative orientation ($\delta\phi$). The black curve follows the potential energy minimum, indicating a bifurcation. (b) and (c) show interaction energy surface cross-sections at two values of $\delta\phi$ and $R$  compared with the classical result (local, $\Lambda=0\text{ \AA}$) .}
    \label{fig:2}
\end{figure*}
\textit{Homologous DNA segments - } The electrostatic interaction between two straight and parallel identical DNA segments of length $L$ at an interaxial separation $R\ll L$ is given by (see Supplemental Material for derivation),
\begin{equation}
    U_{\text{H}}=L\sum_{n=0}^{\infty}a_n(R)\cos(n\delta\phi).
\end{equation}
where $\delta\phi=\phi_1-\phi_2$ is the relative azimuthal orientation of each lateral cross-section of the two molecules, and $\phi_i$ is defined as in Fig. 1. The coefficients $a_n$ are given by: 
\begin{align}
    a_0(R)&=2\pi\bar{\sigma}^2(1-\Theta)^2\mathcal{W}_{0,0}(0,r,a,b),\\
    a_{n\geq1}(R)&=4\pi\bar{\sigma}^2\tilde{p}^2(n)\mathcal{W}_{n,n}(ng,R,a,b),
\end{align}
where $\bar{\sigma}$ is the average intrinsic surface charge density of DNA, $\Theta$ is the percentage of compensated phosphate charge, and,
\begin{equation}
    \tilde{p}(n)=\cos\left(\frac{n\phi_s}{2}\right)-\Theta\left(f_1+f_2(-1)^n+f_3\cos\left(\frac{n\phi_s}{2}\right)\right),
\end{equation}
where $\phi_s\approx 0.8\pi$ is the azimuthal width of the minor groove, $f_1,f_2,f_3$ are the fractions of ions sitting, correspondingly, on the minor groove, major groove and helical strands, and
\begin{widetext}
\begin{align}
    \mathcal{W}_{n,m}(q,R,a,b)=\Omega_{n,m}(q, R, a, a) + \Xi(m,q)\Omega_{n,m}(q,R,a,b)\left(1+\frac{\Xi(-n,-q)}{\Xi(m,q)}+\frac{\Omega_{n,m}(q,R,b,b)}{\Omega_{n,m}(q,R,a,b)}\right),
\end{align}
\begin{align}
    \Omega_{n,m}(q,x,y,z)=4\pi yz\int_0^{\infty}KdK \frac{J_{n-m}(Kx)J_n(Ky)J_m(Kz)}{(K^2+q^2)\varepsilon(\sqrt{K^2+q^2})+\varepsilon\kappa^2},
\end{align}
\end{widetext}
\begin{align}
    \Xi(m,q)=-\frac{a}{b}\frac{\mathcal{A}'_m(a,b,q)-\gamma |q|\frac{I'_m(|q|b)}{I_m(|q|b)}\mathcal{A}_m(a,b,q)}{\mathcal{A}'_m(b,b,q)-\gamma |q|\frac{I'_m(|q|b)}{I_m(|q|b)}\mathcal{A}_m(b,b,q)},
\end{align}
\begin{align}
    \mathcal{A}_m(x,y,q)=4\pi\int_0^\infty KdK \frac{J_m(Kx)J_m(Ky)}{(K^2+q^2)\varepsilon(\sqrt{K^2+q^2})+\varepsilon\kappa^2}.
\end{align}
Above, $J_n(x)$ and $I_n(x)$ are the $n$th order Bessel and modified Bessel functions of the first kind, $\gamma=\varepsilon_c/\varepsilon_*$, and $\mathcal{A}'_m=\partial\mathcal{A}_m/\partial y$.  Closed-form expressions for these integrals are derived in Appendix A of the Supplemental Material. The prefactors $a_n(R)$ sharply decay with $n$, leading to a rapidly-converging alternating sum in Eq. (3), even at small surface-to-surface separations on the order of the Debye length, $(R-2a)\sim\kappa^{-1}$. The nonlocal electrostatic interaction is plotted in Fig. 2 as a function of the interaxial separation $R$ and the relative azimuthal orientation, $\delta\phi$. At close separations a clear enhancement of the interaction energy is seen due to the reduced effective dielectric constant. At intermediate separations, the enhanced screening of the DNA surface charge leads to a ‘crossover’ between the nonlocal and classical models. Interestingly, we found that Eq. (3) is a general result that can also be derived from symmetry considerations, and the coefficients $a_n$ are determined by the specific nonlocal linear response model for the solvent (see Supplemental Material). 

\textit{Helical coherence theory for nonhomologous DNAs - } The specific base-pair sequence of DNA affects the variation in twist angle between adjacent base pairs. This twist angle is on average $\langle\Omega\rangle\approx 35^{\circ}$ with standard deviation $\sigma_{\Omega}\approx 5^{\circ}$ \cite{olson1998}. We modelled the difference in twist angle between two rigid uncorrelated DNA molecules as a random walk with zero mean and $\sqrt{2}\sigma_{\Omega}$ standard deviation~\cite{kornyshev2001}. For long enough molecules the stochastic variable describing the difference in angles, $\delta\phi$, at each cross-sectional cut, is distributed as a Gaussian to a very good approximation. The average interaction energy of two such DNA molecules was therefore calculated as (see Supplemental Material): 
\begin{equation}
    U_{\text{NH}}=L\sum_{n=0}^{\infty}a_n(R)\nu_n(L).
\end{equation}
The coefficients $a_n$ are the same as those defined for homologous DNAs in Eqs. (4) \& (5), and the coefficients $\nu_n$ are a function of the molecule length $L$, but also depend on the following boundary conditions: (i) the point $\ell$ along the DNAs from which mismatches in the sequence start to accumulate, and; (ii) the corresponding value of $\delta\phi$ at that point, denoted by $\xi$. These two parameters are independent as the molecules can freely rotate, and are determined by minimization of Eq. (11) with respect to these parameters. It can be shown (see Supplemental Material) that the optimal values are given by $\ell^*=L/2$, stating that mismatches accumulate from the middle of the molecules which can be understood from symmetry arguments alone, and 
\begin{equation}
    \xi^*=\begin{cases}
        \pm\arccos\left(\dfrac{|a_1|}{a_2}\dfrac{1-e^{-\frac{L}{2\lambda_c}}}{1-e^{-\frac{2L}{\lambda_c}}}\right), \quad \dfrac{|a_1|}{a_2}\dfrac{1-e^{-\frac{L}{2\lambda_c}}}{1-e^{-\frac{2L}{\lambda_c}}}<1\\
        0,\hspace{4.3cm}\dfrac{|a_1|}{a_2}\dfrac{1-e^{-\frac{L}{2\lambda_c}}}{1-e^{-\frac{2L}{\lambda_c}}}>1
    \end{cases}
\end{equation}
which was derived by truncating the sum in Eq. (11) at the second harmonic, $n=2$, possible due to its rapid convergence. The two solutions for $\xi^*$ correspond to regimes of small and large separations respectively. For these optimal parameters, the coefficients $\nu_n$ are given by:
\begin{equation}
    \nu_n(L)=\frac{2\lambda_c}{n^2L}\left(1-e^{-n^2\frac{L}{2\lambda_c}}\right)\cos(n\xi^*)
\end{equation}
where $\lambda_c=h/(\Delta\Omega)^2$ is termed the helical coherence length, and $h\approx3.4\text{ \AA}$ is the vertical rise per base pair. Here, the expression for $\lambda_c$ depends solely on twist angle variations. A more involved expression accounting for different contributions including correlated twist-rise, variation in vertical rise and torsion-stretching elastic thermal fluctuations, has been obtained in the past \cite{kornyshev2013}, resulting in a decreased value of $\lambda_c\approx 100 \text{ \AA}$, also verified experimentally \cite{wynveen2008, kornyshev2013}.

\begin{figure}[t!]
    \centering
    \includegraphics[width=\linewidth]{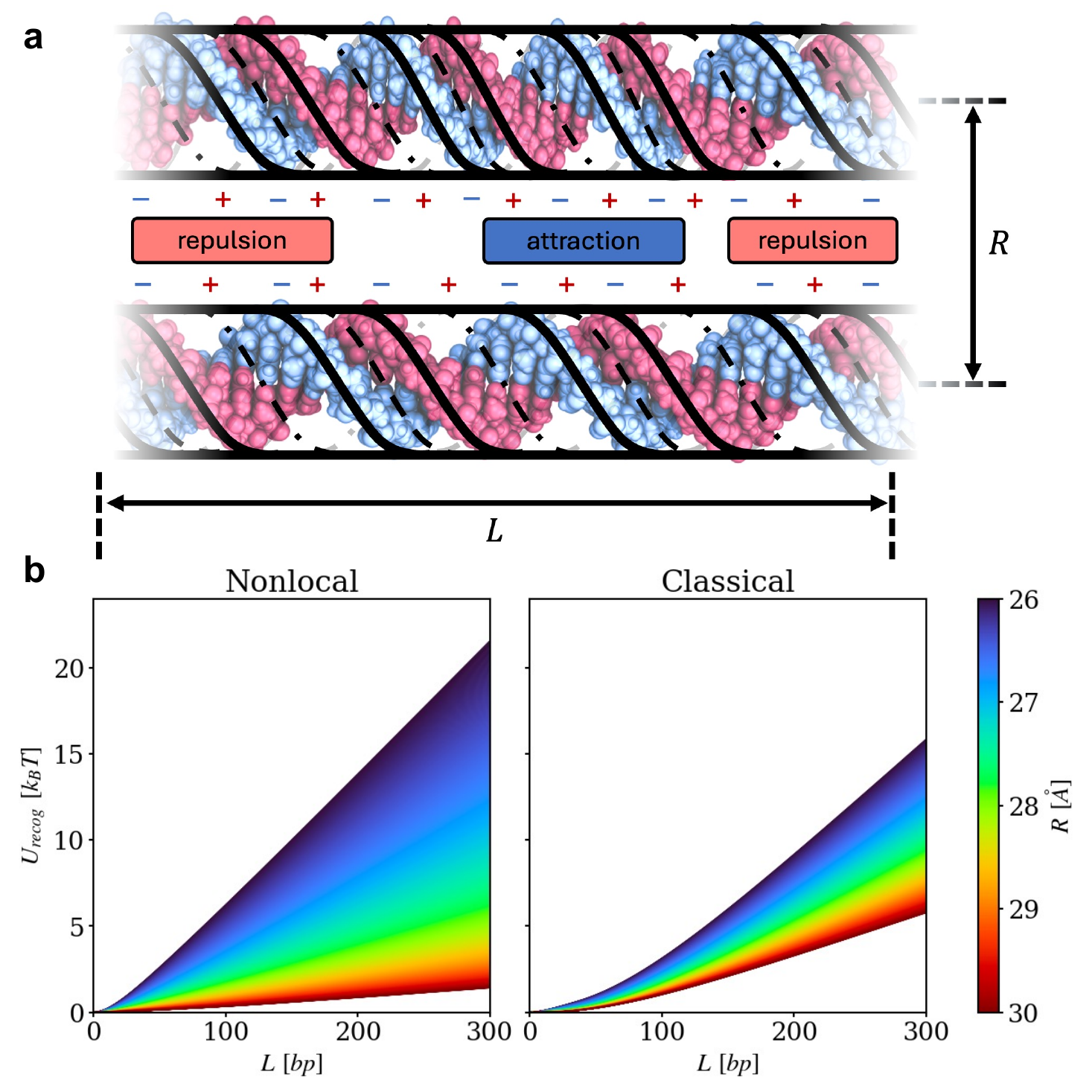}
    \caption{\footnotesize\textbf{DNA recognition energy.} (a) Illustrating how sequence-dependent twist angle variations in non-homologous DNA lead to both electrostatically attractive and repulsive interacting segments. (b) shows variation of the recognition energy as a function of DNA length for a range of interaxial separations, comparing between nonlocal and classical models of the solvent dielectric response. All system parameters are the same as in Fig. 2.}
    \label{fig:3}
\end{figure}
\begin{figure*}
    \centering
    \includegraphics[width=\linewidth]{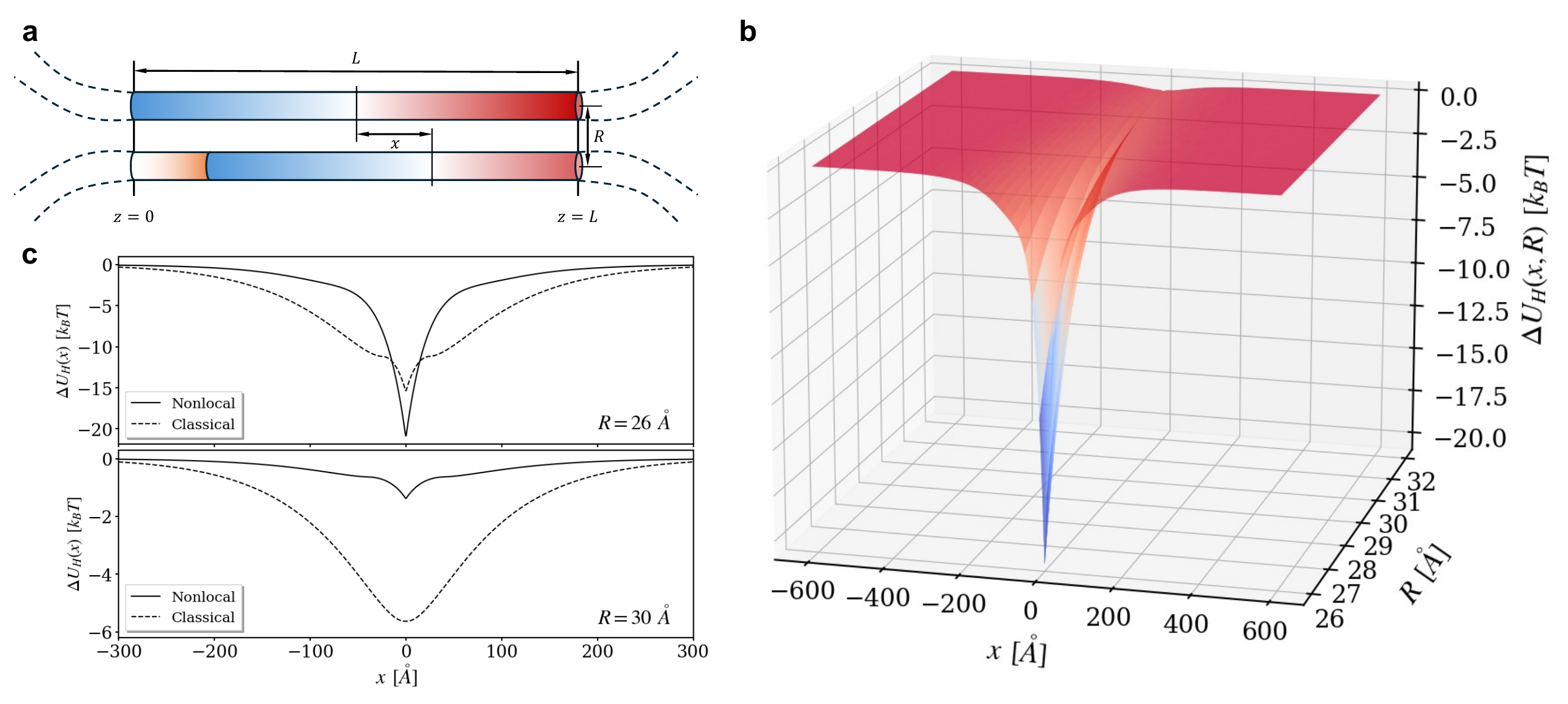}
    \caption{\footnotesize\textbf{Homology recognition well.} (a) An illustration of two parallel homologous DNAs, displaced by a distance $x$, within a juxtaposition window of a single Kuhn length, $L=1000\text{ \AA}$. Sequence homology is indicated by the red/blue gradient. The section of DNA that leaves the juxtaposition window at $z=L$ is replaced by a new section at $z=0$. (b) Recognition well as a function of the interaxial separation, $R$ and the axial shift, $x$. (c) shows cross-sections of the surface below and above the crossover point.}
    \label{fig:4}
\end{figure*}
\textit{Recognition energy and homology recognition well - } A direct indication of the recognition ability of homologs over non-homologs is given by the recognition energy, $U_\text{recog}$, defined as the difference in interaction energies of homologs (Eq. (3)) and non-homologs (Eq. (11)): 
\begin{equation}
    U_{\text{recog}}=U_{\text{NH}}-U_{\text{H}}.
\end{equation}
Comparing $U_{\text{recog}}$ for nonlocal against local ($\Lambda=0$ or $\varepsilon_*=\varepsilon$) electrostatic models of the solvent (Fig. 3) shows (i) a significant enhancement at small separations, and (ii) a reduction at larger separations, with some crossover in between. The former follows from a reduced effective dielectric constant near the DNA surface in the nonlocal model. The latter effect arises at larger DNA separations from the enhanced screening by the electrolyte ions, which will also interact within the same reduced effective dielectric constant. The crossover separation depends on the chosen coupling between ionic and solvent dielectric responses. However, the described qualitative behaviour is expected to remain independent of the coupling model. At very large separations, the models coincide.

Another important feature of homology recognition is the work $U_\text{H}(x)$ required to shift two parallel homologs, initially in perfect register, by a displacement $x$ along their lengths, within a juxtaposition window of length $L$ (Fig. 4a) \cite{kornyshev2009}. This energy is given by (see Supplemental Material): 
\begin{equation}
    U_{\text{H}}(x)=L\sum_{n=0}^\infty a_n(R)\mu_n(x,L)\cos(n\xi^*(x,L)).
\end{equation}
where
\begin{equation}
    \mu_n(x,L)=\begin{cases}
        \dfrac{2\lambda}{n^2L}\left(1-e^{-\frac{n^2 x}{\lambda_c}}\right)+\left(1-\frac{2x}{L}\right)e^{-\frac{n^2x}{\lambda_c}},\hspace{0.2cm}x\leq \frac{L}{2}\\\\
        \dfrac{2\lambda_c}{n^2L}\left(1-e^{-\frac{n^2L}{2\lambda_c}}\right),\hspace{2.8cm}x>\frac{L}{2}
    \end{cases}
\end{equation}
and
\begin{equation}
    \xi^*(x,L)=\begin{cases}
        \pm\arccos\left(\frac{a_1 \mu_1(x,L)}{4a_2\mu_2(x,L)}\right),\quad\quad \left|\frac{a_1\mu_1(x,L)}{4a_2\mu_2(x,L)}\right|\leq 1\\\\
        0,\hspace{3.8cm}\left|\frac{a_1\mu_1(x,L)}{4a_2\mu_2(x,L)}\right|>1
    \end{cases}.
\end{equation}
Eqs. (16) and (17) were determined by minimizing the energy with respect to the two boundary parameters; the point $\ell$ from which mismatches accumulate and $\xi$, the difference in azimuthal angles at that point. 

Noticing that a shift by $x=L/2$ corresponds, effectively, to a non-homologous pair, we plotted $\Delta U_\text{H}(x)=U_\text{H}(x)-U_\text{H} (L/2)$ in Fig. 4 to compare nonlocal and local models for various DNA-DNA separations. The profile of $\Delta U_\text{H}(x)$ is a funnel-shaped well with a minimum at $x=0$, where the homologs are at complete register. Hence, the depth of the well corresponds to the recognition energy, $U_{\text{recog}}$. For the nonlocal electrostatic model, homologs which diffuse in close proximity to each other simply click into register due to the extreme deepness and sharpness of the recognition funnel. 

Thus, it transpires that the correlations of dipole moment fluctuations in water dramatically enhance the recognition energy, deepening and sharpening the recognition funnel that draws the two homologs into an ideal \textit{similia-similibus} juxtaposition \cite{falaschi2008}. This finding substantiates the hypothesis of an electrostatic mechanism, or at least one important factor, that favours recognition of homologous genes prior to the assistance of any proteins.
\begin{acknowledgements}
We are thankful to Yuri Budkov, Roland Kjellander and H\'el\`ene Berthoumieux for useful discussions. This work is supported by the Leverhulme Trust grant \#RPG-2022-142 (EH and AAK) and the Imperial College President's PhD Scholarship (JGH).
\end{acknowledgements}
\bibliography{DNA_hom_rec}
\end{document}